\documentclass[final,5p,times,twocolumn]{elsarticle}

\usepackage{graphicx}
\usepackage{amssymb}
\usepackage{amsmath}
\journal{Physica E}

\begin{document}

\begin{frontmatter}

%% Title, authors and addresses

%% use the tnoteref command within \title for footnotes;
%% use the tnotetext command for theassociated footnote;
%% use the fnref command within \author or \address for footnotes;
%% use the fntext command for theassociated footnote;
%% use the corref command within \author for corresponding author footnotes;
%% use the cortext command for theassociated footnote;
%% use the ead command for the email address,
%% and the form \ead[url] for the home page:
%% \title{Title\tnoteref{label1}}
%% \tnotetext[label1]{}
%% \author{Name\corref{cor1}\fnref{label2}}
%% \ead{email address}
%% \ead[url]{home page}
%% \fntext[label2]{}
%% \cortext[cor1]{}
%% \address{Address\fnref{label3}}
%% \fntext[label3]{}

\title{Control of a two-electron quantum ring with
an external magnetic field}

%% use optional labels to link authors explicitly to addresses:
%% \author[label1,label2]{}
%% \address[label1]{}
%% \address[label2]{}

\author{J.~S\"arkk\"a \corref{cor1}} 
\ead {Jani.Sarkka@tkk.fi} 
\author{A.~Harju}
\cortext[cor1]{Corresponding author. Tel.:+358 9 451 5157; fax: +358 9 451 3241.}
 
\address{
Helsinki Institute of Physics and Department of Applied Physics, \\ Helsinki University of Technology, P.O.Box
4100, FI-02015 TKK, Finland}

\begin{abstract}
%% Text of abstract
We investigate the use of external time-dependent magnetic
field for the control of the quantum states in
a two-electron quantum ring.
The hyperfine interaction of the confined electrons with
surrounding nuclei couples the singlet state 
with the three triplet states. 
When the external magnetic field is changed, the singlet
ground state becomes degenerate with the triplet
states allowing singlet-triplet transitions.
By choosing different speeds for the magnetic
field switching the final quantum state
of the system can be manipulated. 
We evaluate suitable magnetic field values and time scales for
efficient quantum ring control.
\end{abstract}

\begin{keyword}
%% keywords here, in the form: keyword \sep keyword
Quantum dots \sep Quantum computation
%% PACS codes here, in the form: \PACS code \sep code
\PACS 73.21.La \sep 42.50.Dv \sep 71.70.Gm \sep 71.70.Jp
%% MSC codes here, in the form: \MSC code \sep code
%% or \MSC[2008] code \sep code (2000 is the default)
\end{keyword}

\end{frontmatter}

%% \linenumbers

%% main text
\section{Introduction}
\label{}

The construction of a working quantum computer is
one of the most fascinating aims of modern science.
The building block of a quantum computer is
qubit, a quantum bit. During recent years, several proposals
for a qubit have been studied experimentally and theoretically.
One of the most popular candidates has been electron spin in
semiconductor nanostructures.

The control of a two-electron double quantum dot has been realized
experimentally by using electric fields \cite{petta05}
and has also been studied theoretically \cite{murgida:036806,sarkka:245315}.
When the electron density is concentrated over a certain radial
distance from the dot center, the quantum dot is transformed 
to a quantum ring.
The tuning of the quantum states of quantum rings by electric
fields has been demonstrated experimentally \cite{fuhrer:206802},
the energy structure of two-electron quantum rings
has been focus of theoretical investigations 
\cite{climente2005,malet:1492,xie2008},
and feasible qubit operations for the quantum ring 
have also been discussed \cite{waltersson:115318}.
The decoherence caused by the interaction between the confined electron
spin and the hyperfine spins of surrounding nuclei
has been widely studied \cite{deng:241303,taylor:035315,nepstad:125315}.
Recently, the control of two-electron double quantum dot
by time-dependent external magnetic field has been
addressed \cite{sarkka:045323}. Here we apply this
control method in a quantum ring.
In our analysis, the electrons are
confined by a harmonic radial potential.
In addition, there is a Gaussian potential
at the dot center. This causes the
electron density to concentrate in a ring-shaped
area around the origin, making a quantum ring.

In this paper, we study the control of a two-electron
quantum ring using magnetic field. The lowest-lying energy
states of this system are the singlet state $|S \rangle$
and three triplet states $|T_{-} \rangle$, $|T_{0} \rangle$,  
and $|T_{+} \rangle$. The hyperfine interaction couples the singlet and triplet
states. When the singlet and triplet states are degenerate,
singlet-triplet transitions are possible. We show that the
time-dependent external magnetic field can be used to control
the singlet and triplet energies in order to induce
transitions from singlet to triplet state.

The rest of the manuscript is organized as follows.
The physical model is discussed in Section 2.
In Section 3 the numerical results are presented
and conclusions are made in Section 4.

\section{Model}
\begin{figure}
\begin{center}
\includegraphics[width=0.39\textwidth]{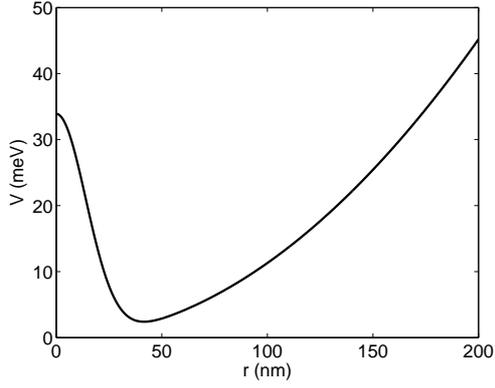}
\end{center}
\caption{
\label{fig:potential}
The potential of the quantum ring,
given by Eq.~(\ref{eq:ringpot})}
\end{figure}

We model the two-electron system with the Hamiltonian
\begin{equation}
H=\sum_{i=1}^{2} \Bigg(\frac{\Big(-i\hbar \nabla_{i}
-\frac{e}{c}\mathbf{A}_{i}\Big)^{2}}{2m^{*}}
+V(\mathbf{r}_{i},\mathbf{s}_{i}) \Bigg)
+\frac{e^{2}}{\epsilon r_{12}},
\end{equation}
where the effective mass $m^*$=$0.067m_e$ 
and permeability $\epsilon$=12.7
are material parameters for GaAs.
The external potential $V$ consists of
two parts,
\begin{equation}
V=V_{Z}+V_{C}.
\end{equation}
The first part, $V_{Z}$, is the potential caused by the 
Zeeman interaction
\begin{equation}
V_{Z}(\mathbf{r},\mathbf{s})=g^{*}\mu_{B}\mathbf{B}(\mathbf{r})\cdot
\mathbf{s},
\end{equation}
where the Land\'e factor of GaAs is $g^{*}$=-0.44.
The magnetic field can be divided into a homogeneous 
external magnetic field $\mathbf{B}_{ext}=\nabla \times \mathbf{A}$ and an inhomogeneous random
hyperfine field $\mathbf{h}(\mathbf{r})$.
The hyperfine field strength is determined
from a fit to experimental data \cite{sarkka:245315}.
The second part in the external potential is 
the confinement potential $V_{C}$, that depends only on the
radial coordinate. In our case it has two parts, harmonic potential
and Gaussian peak at the center of the ring, resulting
to the expression \cite{gustav}
\begin{equation}
\label{eq:ringpot}
V_{C}(r)=\frac{1}{2}m^{*}\omega_{0}^{2}r^2
+V_0 \exp(-r^2/{r_0}^2),
\end{equation}
where the Gaussian peak has
height $V_0$=34 meV, width $r_0$=20 nm, and the
the strength of the harmonic potential is $\hbar \omega_{0}$=3.0 meV.
The radial potential is illustrated in Fig.~\ref{fig:potential}.

\begin{figure}[hb]
\begin{center}
\includegraphics[width=0.49\textwidth]{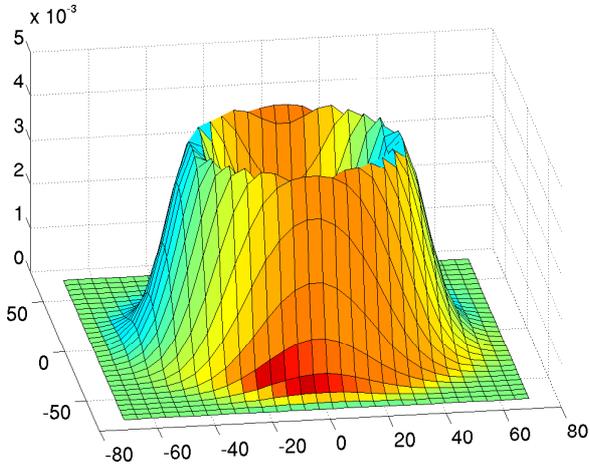}
\end{center}
\caption{
\label{fig:density}
Electron density of the singlet state
in a quantum ring. Unit of the $x$ and $y$
coordinates is nm. Magnetic field strength is 1.3 T.}
\end{figure}

We solve the lowest-energy singlet and
triplet states using finite difference method
with Lanczos diagonalization. 
The density of the singlet state of quantum
ring is depicted in Fig.~\ref{fig:density}. The maximum of
the singlet density is attained at the radius
$r \approx 40$ nm, where the confining potential has 
its minimum value.
The Gaussian peak added to the confinement 
potential at the center of the dot creates
a hole in the singlet density of the quantum ring.
The singlet density vanishes for radii
larger than 60 nm.

\begin{figure}
\begin{center}
\includegraphics[width=0.39\textwidth]{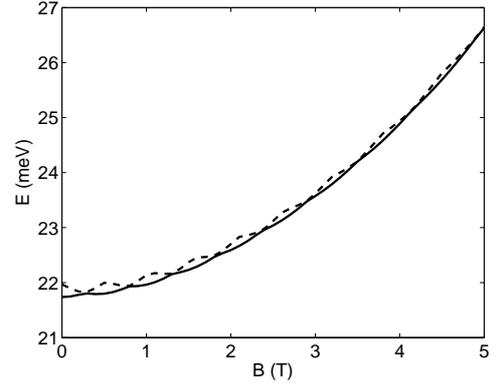}
\end{center}
\caption{
\label{fig:energy}
Energies of the ground state (solid) and first excited state
(dashed) of the quantum ring as a function of the external
magnetic field. In the calculation of
these energies, the hyperfine field is set to zero and
the Zeeman term is neglected.}
\end{figure}

The energies of the two lowest energy states as
a function of the external magnetic field
are shown in Fig.~\ref{fig:energy}. In the
calculation, the hyperfine field and the
Zeeman term in the Hamiltonian are not taken
into account, because these terms bring
at most 0.1 meV contribution to the total energy
for magnetic fields smaller than 5 T.
Hence, the two lowest states are singlet state
and triplet state. For zero magnetic field 
singlet is the ground state. As the magnetic
field increases, singlet and triplet states
alternate in ground state and excited state.

\begin{figure}
\begin{center}
\includegraphics[width=0.39\textwidth]{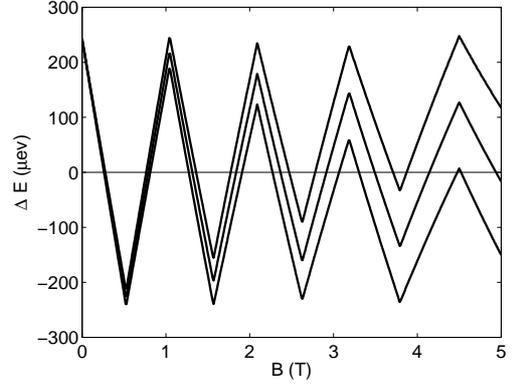}
\end{center}
\caption{
\label{fig:enedif}
The energy difference $\Delta E$ between the singlet state
$S$ and triplet states $T_-,T_0$ and $T_+$ of a double
quantum dot as a function
of the external magnetic field. In the calculation of
these energies, the hyperfine field is set to zero.} 
\end{figure}

The singlet-triplet transitions are possible
only for such magnetic field values when
the singlet and triplet become degenerate.
Thus, the relevant energy for the following
analysis is the difference $\Delta E=E_T-E_S$
between singlet and triplet energies $E_S$ and $E_T$.
The energy differences between
the three triplet states and the singlet state
as a function of magnetic field are shown in Fig.~\ref{fig:enedif}.  
Here the Zeeman term is included in the Hamiltonian,
but hyperfine term is omitted, as it does not
contribute significantly to the energy difference.
The mechanism of the quantum ring control is
based on the fact that the energy differences change
sign when the magnetic field increases.
When the magnetic field is increased,
the singlet and triplet states are degenerate for certain 
magnetic field values. The degeneracies of the
three triplet states occur at differing magnetic fields,
and the differences between the degeneracy points
increase for larger magnetic fields due to the 
Zeeman energy.

There are several singlet-triplet degeneracies for
different magnetic field values, enabling the control of
the speed of the singlet-triplet transition not
only by changing the sweeping time of the magnetic field,
but also by choosing different magnetic field intervals,
as the energy changes more slowly as a function of magnetic
field for larger magnetic field values and the three
degeneracy points are further apart from each other.
If the confining potential is stronger than the one used
in our analysis, the maximum of the energy difference
increases. In such case, the singlet-triplet degeneracies
are closer to each other. Hence, the interval of
the magnetic field values used in the switching would
need to be much smaller.
%The dependence of the energy difference on the magnetic
%field is rather linear.
%Due to this linear dependence, the singlet-triplet 
%transition probabilities are given exactly by the Landau-Zener 
%formula \cite{landau32,zener32}.

The energy difference $\Delta E$ depends linearly on the magnetic
field under many intervals. If the magnetic field is changed linearly as a 
function of time inside such an interval, $\Delta E$ depends
linearly on time. If $\Delta E $ changes sign in the interval,
the occurring transition is a Landau-Zener transition.
For a two-level quantum system, the probability of a
diabatic transition depends only on the off-diagonal 
element of the two-level Hamiltonian $\alpha$, and on 
the time derivative of the energy difference $K$,
according to the Landau-Zener formula  \cite{landau32,zener32}
\begin{equation}
P=\exp\Bigg(-\frac{2 \pi \alpha^{2}}{|K|}\Bigg).
\end{equation}
If the hyperfine field coupling the two states is random,
one has to integrate the probability over Gaussian
distributed $\alpha$, resulting to a more complicated expression \cite{sarkka:045323}.

In our setup, the energies of the states above
the four lowest-lying states are considerably larger than their coupling
with the lowest-lying states induced by the hyperfine field.
Hence, we approximate the quantum dynamics of the system using a
4 $\times$ 4 Hamiltonian, constructed in the basis of 
the singlet and three triplet states \cite{coish:125337}.
The system is initialized in the singlet states.
The time dependence of the quantum states is obtained
from the relation $\psi(t)=\exp(-iHt/\hbar)\psi(0)$.
The randomness of the hyperfine field is taken into 
account by averaging $\psi(t)$ over 1000 different realizations
of the hyperfine field. The probabilities of the singlet and
triplet states are obtained from the squares of the
expansion coefficients of the wave function in the singlet-triplet basis.

\begin{figure*}[ht]
\begin{center}
\includegraphics[width=0.39\textwidth]{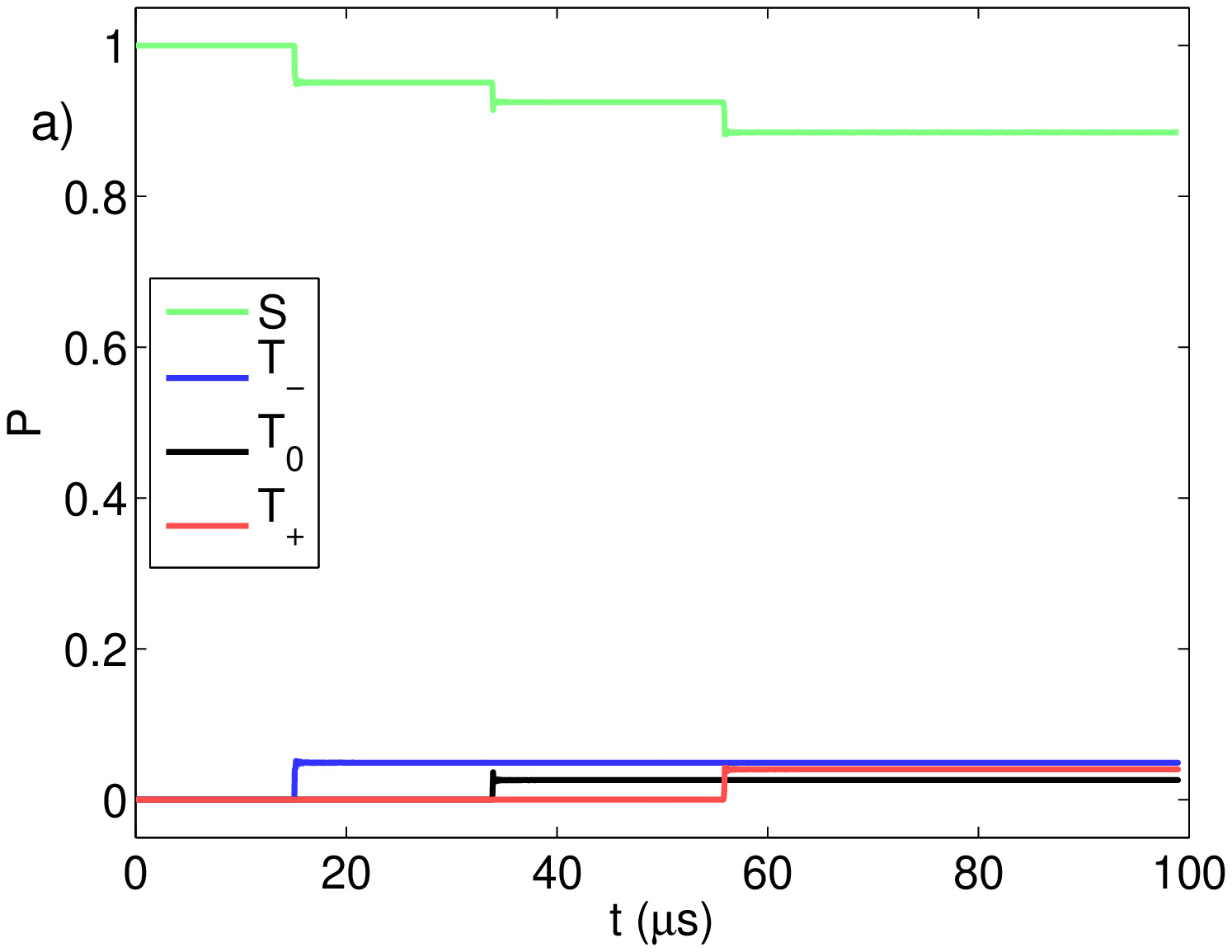}
\includegraphics[width=0.39\textwidth]{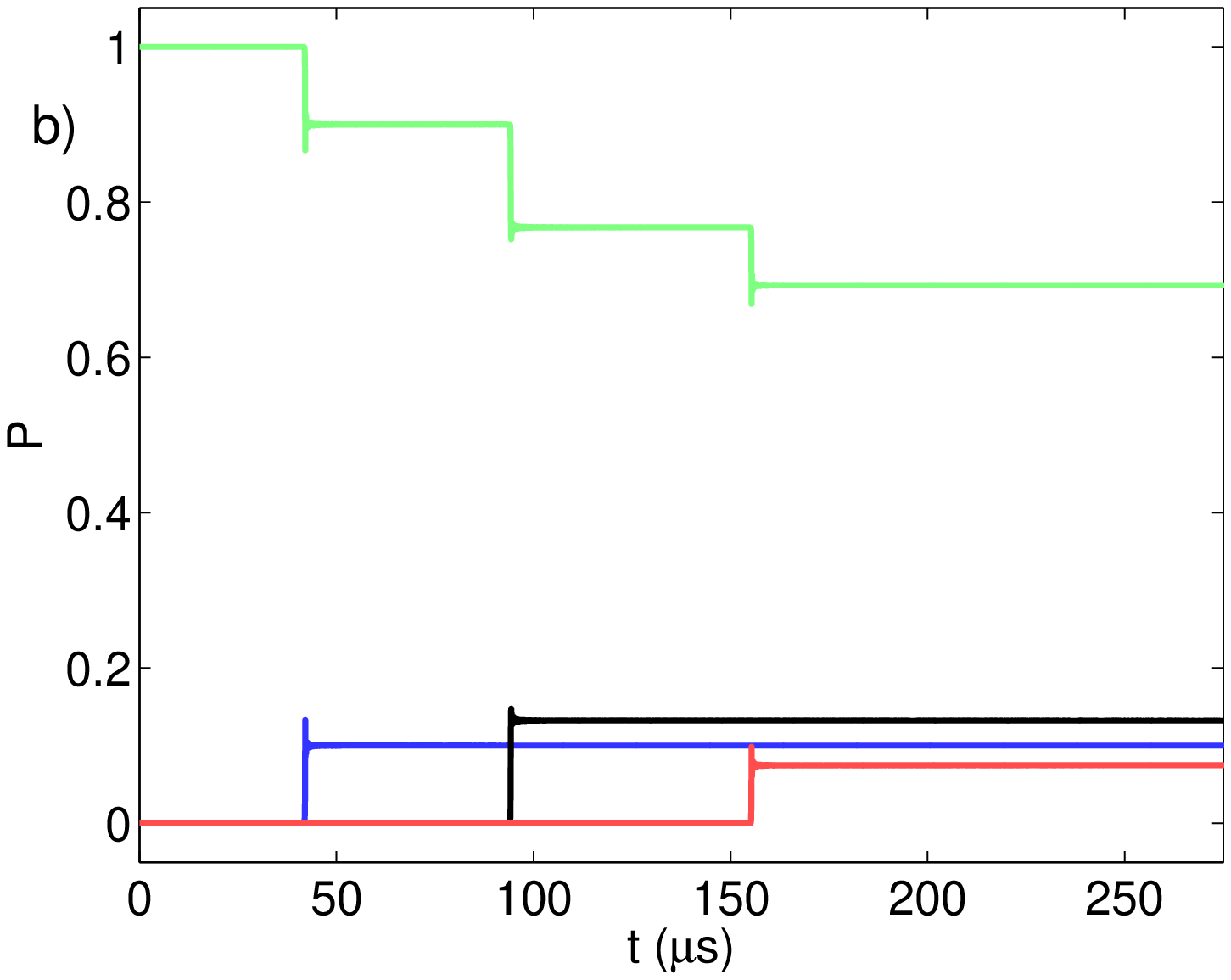}\\
\includegraphics[width=0.39\textwidth]{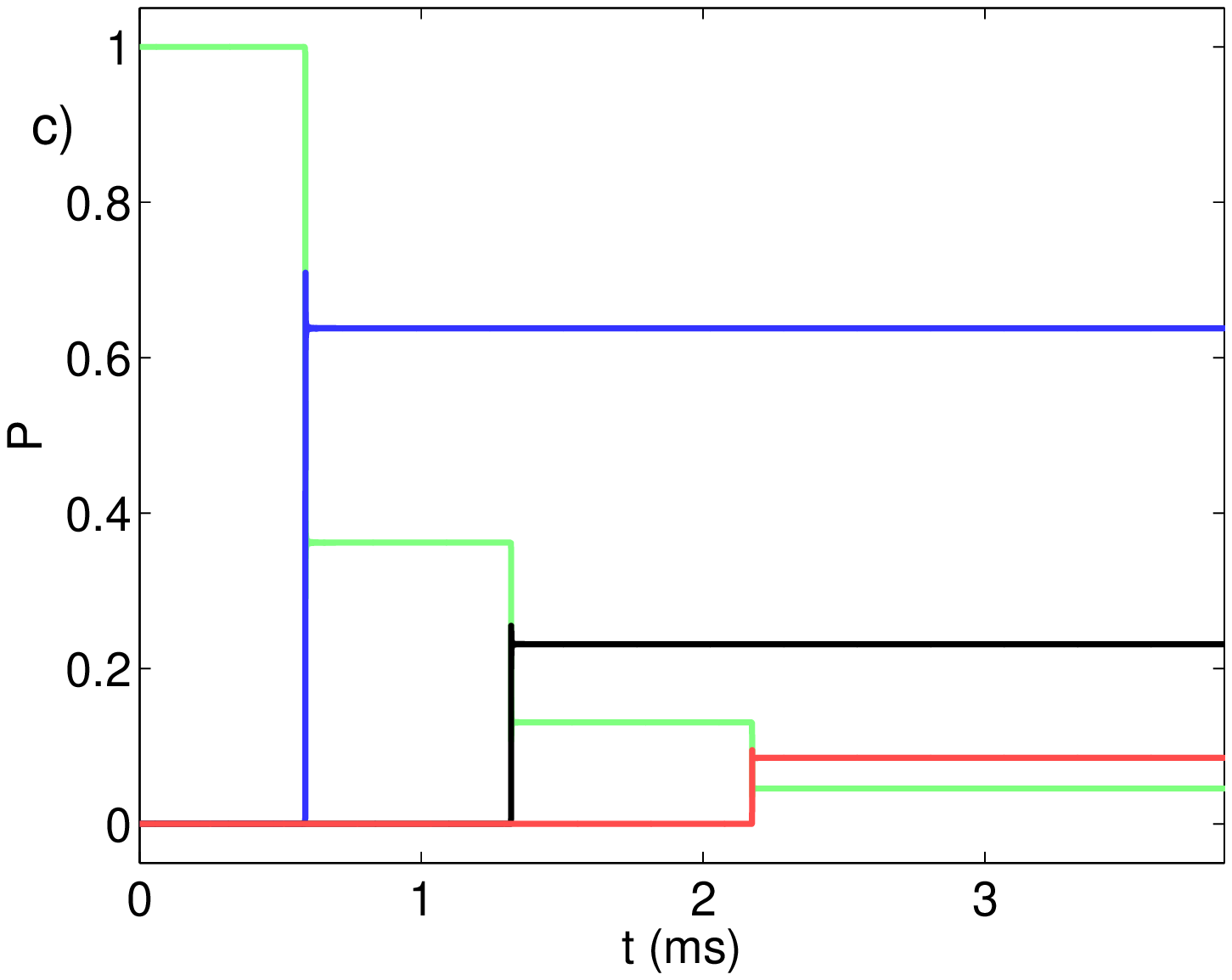}
\includegraphics[width=0.39\textwidth]{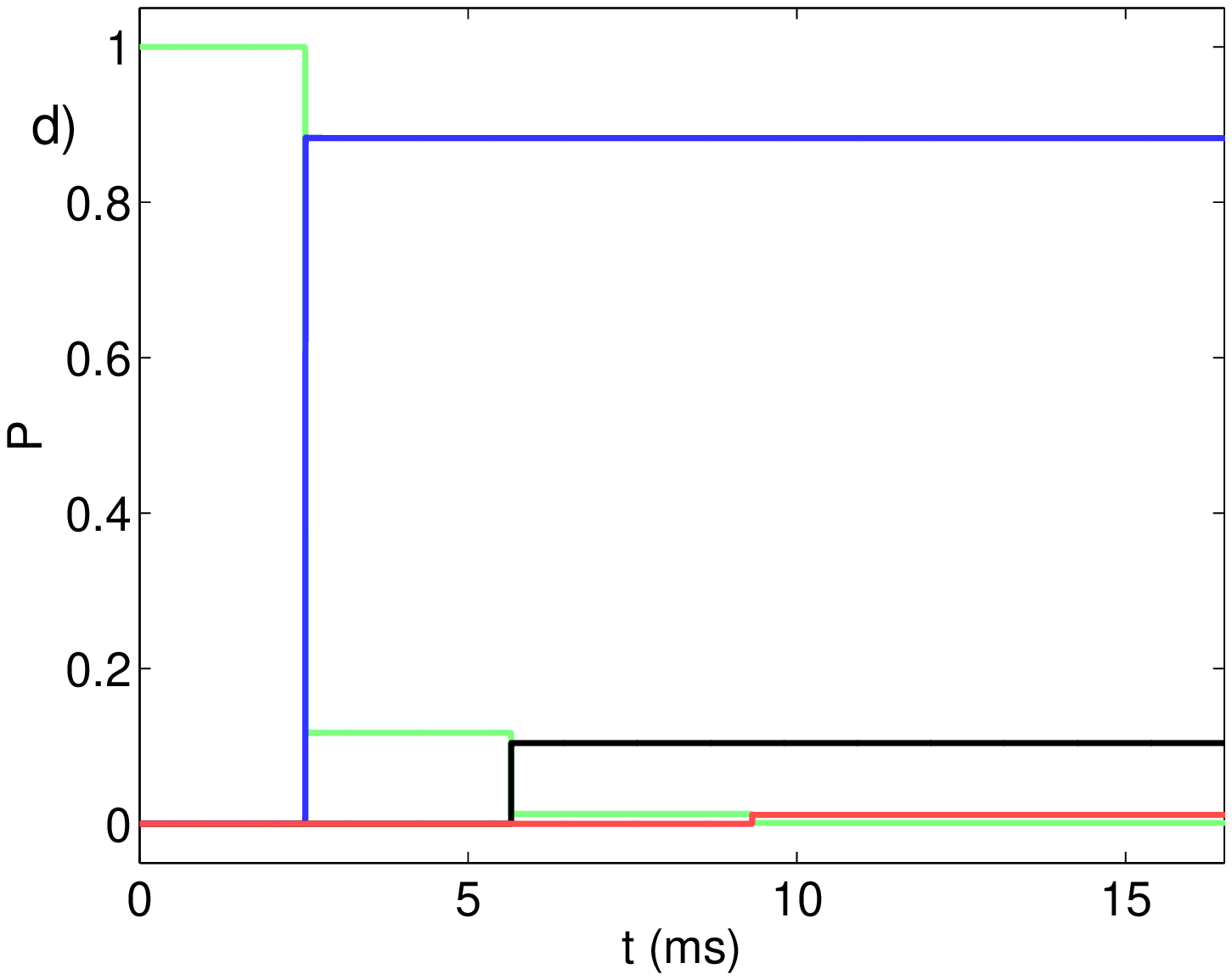}
\end{center}
\caption{
\label{fig:steps}
Probabilities $P$ of the
singlet state $S$ and triplet 
states  $T_-,T_0$ and $T_+$ of a quantum ring
as a function of time averaged over
1000 realizations. 
The time used to switch the magnetic field from
1.2 T to 1.5 T is 100 microseconds in (a),
275 microseconds in (b), 3.85 milliseconds in (c),
and 17 milliseconds in (d).}
\end{figure*}

\section{Results}

In the control scheme we use, the magnetic field is changed 
linearly from 1.2 T to 1.5 T. All three triplet states have degeneracies with
the singlet state inside this interval, see Fig.~\ref{fig:enedif}.
Hence, the singlet probability decreases three times, resulting
in a graph where the singlet probability has a step function form with
three steps. In Fig.~\ref{fig:steps} are shown the probabilities of the
singlet and triplet states as a function of time,
averaged over 1000 random hyperfine field realizations, for four
different magnetic field switching times: 100 $\mu$s (a), 275 $\mu$s (b),
3.85 ms (c), and 17 ms (d). All four figures indicate that the
singlet-triplet transitions occur at the close neighborhood
of the degeneracy points, as the probabilities stay constant 
until the degeneracy point is reached.
The oscillations of singlet and triplet states are so rapidly damped 
that they are not visible in Fig.~\ref{fig:steps}. Only peaks at
the transition points remain as traces of the oscillations.
Hence, the behavior of the probabilities resembles step function.
In Fig.~\ref{fig:steps} (a), the switching is done in 100 $\mu$s.
Now the switching is so fast that the singlet probability
does not change considerably at the degeneracy points, hence the triplet
probabilities are quite small in the end. When the switching takes
275 $\mu$s in Fig.~\ref{fig:steps} (b), we observe that the steps 
of the singlet probability are now larger, but still the singlet probability is
in the end larger than the triplet probabilities. If the switching time
is increased to 3.85 ms, see Fig.~\ref{fig:steps} (c),
we denote that the singlet probability changes considerably
already at the first singlet-triplet transition. The
following two transitions are thus smaller, as the singlet 
probability has a diminished value before the transitions take place.
Finally in the Fig.~\ref{fig:steps} (d), the switching time is increased
to 17 ms. Now the slow switching process enables a large singlet-triplet
transition at the first degeneracy point and $T_+$ state gains
a probability close to 1. Thus, the other triplet states can only
have small probabilities in the end.
These differences in these four figures are caused by
the exponential dependence of the singlet-triplet transition 
probability on the switching speed.
For a single hyperfine field realization,
the behavior of the singlet probability gives a qualitatively similar step
function graph, where the transition probabilities are given exactly
by the Landau-Zener formula.
% $\exp(-2\pi D/|K|)$, 
%where $D$ is proportional to the hyperfine field strength
%and $K$ is proportional to the switching speed.
%The Landau-Zener formula is modified by the integration over
%different hyperfine field values \cite{sarkka:045323}.
For slower switching speeds in Figs.~ \ref{fig:steps} (c) and (d)
the $T_-$ state has a large probability after the transition.
This limits the probabilities of the $T_0$ and $T_+$ states
due to the conservation of the total probability.

\section{Conclusions}
In conclusion, we analyzed the use of an external
time-dependent magnetic field for the control
of a two-electron quantum ring. 
The energy structure of the singlet and triplet
states allows singlet-triplet transitions
for certain magnetic field values. By 
initializing the system in the singlet state and
changing the external magnetic field value continuously,
so that singlet-triplet degeneracy points are crossed,
all three triplet states have nonzero probabilities.
By choosing
different switching times for the magnetic field,
one obtains a variety of final states,
as they are different superpositions
of the singlet and three triplet states.
In the present analysis, the speed of
the magnetic field change was constant
during the switching process.
This speed could be changed between different
singlet-triplet transitions. Thus,
the probabilities of the different triplet
states can be easily adjusted, making this
setup feasible for the construction of a qubit.
Our results give the relevant time scales
and magnetic field values for the
experimental realization of quantum ring control.
The singlet-triplet energy behaves piecewise
linearly as a function of the external magnetic field.
The Landau-Zener formula, valid for linear energy crossings,
can be used for accurate description
of singlet-triplet transition processes.
The time scale of the magnetic field
change from 1.2 T to 1.5 T in this control
scheme is of the order of 0.1 -1 ms. For
the time being, such a magnetic field control
is quite demanding to do experimentally,
but the development of experimental techniques
may solve this problem in the near future.

\section*{Acknowledgment}

This study has been supported by the Academy of Finland through
its Centers of Excellence Program (2006-2011).

%% The Appendices part is started with the command \appendix;
%% appendix sections are then done as normal sections
%% \appendix

%% \section{}
%% \label{}

%%\begin{thebibliography}{eparticle}

%% \bibitem{label}
%% Text of bibliographic item

%% \bibitem{}
%% \bibliography{eparesub}

\end{document}